\documentclass[12pt,preprint]{aastex}

\shorttitle{Jet Sideways Expansion Effect on Estimating the
Gamma-Ray Burst Efficiency} \shortauthors{Zhao & Bai}

\begin{document}


\title{Jet Sideways Expansion Effect on Estimating the Gamma-Ray Burst Efficiency
   }


\author{Xiaohong Zhao\altaffilmark{1,2}}

\author{J. M. Bai\altaffilmark{1}}
\altaffiltext{1}{National Astronomical Observatories/Yunnan
Observatory, Chinese Academy of Sciences, P.O. Box 110, 650011
Kunming, China; zhaoxh@vip.sohu.com.} \altaffiltext{2} {Graduate
School of the Chinese Academy of Sciences, 100012, Beijing, China.}


\begin{abstract}
The high efficiency of converting kinetic energy into gamma-rays
estimated with late-time afterglows in Gamma-Ray Burst (GRB)
phenomenon challenges the commonly accepted internal-shock model.
However, the efficiency is still highly uncertain because it is
sensitive to many effects. In this Letter we study the sideways
expansion effect of jets on estimating the efficiency. We find that
this effect is considerable, reducing the efficiency by a factor of
$\sim0.5$ for typical parameters, when the afterglow data $\sim 10$
hr after the GRB trigger are used to derive the kinetic energy. For
a more dense circumburst medium, this effect is more significant. As
samples, taking this effect into account, we specifically calculate
the efficiency of two bursts whose parameters were well constrained.
Almost the same results are derived. This suggests that the sideways
expansion effect should be considered when the GRB efficiency is
estimated with the late afterglow data.

\end{abstract}


\keywords{gamma ray: bursts --- gamma ray: observations}



\section{Introduction}
The efficiency ($\eta_\gamma$) of converting kinetic energy into
$\gamma$-rays ($E_\gamma$) is one of the most important physical
parameters of gamma-ray burst (GRB) phenomenon (Zhang et al. 2007,
hereafter Z07). In the conventional internal+external shock GRB
model (Rees \& M\'{e}sz\'{a}ros 1994; Kobayashi, Piran \& Sari 1997;
see recent review by Zhang \& M\'{e}sz\'{a}ros 2004), $\eta_\gamma$
is $\sim 1\%-5\%$ (e.g., Kobayashi, Piran \& Sari 1997; Kumar 1999),
and even under some extreme assumptions, such as the extremely
inhomogeneous velocity of the ejecta shells, $\eta_\gamma$ is only
40\% (Kobayashi, Piran \& Sari 1997, Kobayashi \& Sari 2001).
However, the $\eta_\gamma$ estimated with the late afterglow (say,
10 hr after the GRB trigger) are much higher ($>50\%$) than the
model prediction (Lloyd-Ronning \& Zhang 2004, hereafter LZ04). By
considering the inverse Compton effect, Fan \& Piran (2006,
hereafter FP06) derived $\eta_\gamma=1\% \sim 89\%$. More recently,
Z07 revisited this issue for 32 GRBs detected by \emph{Swift} with
the early X-ray afterglow data. They derived the kinetic energy
(isotropic-equivalent) at the end of the shallow decay phase ($t_b$)
and at the deceleration time of the fireball ($t_{dec}$), and they
found that $\eta_\gamma$ estimated with the kinetic energy at $t_b$
is $<10\%$ for most of the GRBs and that the $\eta_\gamma$ estimated
with the kinetic energy at $t_{dec}$ varies from a few percent to
$>90\%$. These results challenge the conventional internal-shock
model.

The difficulty of measuring $\eta_\gamma$ is the derivation of the
kinetic energy ($E_k$) of the afterglows, which sensitively depends
on many physical parameters of the burst itself and its environment
(Z07). Measurement of the $E_k$ with early afterglow data seems more
reliable (Z07). However, it is difficult to determine the onset of
afterglow from the observation. Furthermore, most of the X-ray early
afterglows detected by \emph{Swift} show complex features with
irregular flares (Zhang et al. 2006; Nousek et al. 2006; Panaitescu
et al. 2006; Granot et al. 2006). Estimating the kinetic energy with
them can also introduce some uncertainties. Zhang et al. (2007)
systematically analyzed the prompt and early afterglow emission of
32 \emph{Swift} bursts and found that the GRB efficiency derived
with early afterglow is highly variable. This result---if not
intrinsic---seems to prove the above viewpoint. The derivation of
$E_k$ with the late afterglow data, on the other hand, involves some
uncertainties due to the energy loss, energy injection, and jet
sideways expansion effect.  The advantage of the late afterglow is
that this phase is steady, generally showing a smooth power-law
decay. By taking the energy loss, the injection, and the jet
expansion effect into account, the late afterglow may also be used
for estimating $\eta_\gamma$.

It is now believed that GRBs arise from jets collimated into a small
angle. To derive the actual radiative and kinetic energy from the
observed fluence, the collimation correction has to be considered.
In previous works, the isotropic-equivalent $E_{\gamma}$ and $E_{k}$
were used to derive the efficiency without considering this
geometrical correction, which is only valid when the effect of the
jet expanding sideways is ignored. However, the GRB jet evolves and
expands sideways with time. Although this effect is not notable at a
very early stage (Huang et al. 2000), it may affect the estimation
of the efficiency $\eta_\gamma$, when one uses late afterglows at
$\sim 10$ hr to calculate the efficiency. In this {\em Letter}, we
investigate the jet sideways expansion effect on estimating the GRB
efficiency. Our theoretical analysis is present in \S 2. Cases
studies for some typical GRBs are present in \S 3. Discussion and
conclusion are presented in \S 4.

\section{Theoretical Analysis}
The $\eta_\gamma$ is defined as
\begin{equation}
\eta_\gamma\equiv \frac{E_{\gamma}}{E_{k}+E_{\gamma}}.
\end{equation}
Assuming $f_{b,\gamma}$ and $f_{b,k}$ are the beaming factors during
the prompt and the afterglow phases, respectively, the corrected
efficiency $\eta_\gamma^c$ is given by

\begin{equation}
\eta_\gamma^{c}=
\frac{E_{\gamma}f_{b,\gamma}}{E_{k}f_{b,k}+E_{\gamma}f_{b,\gamma}}=\frac{\eta_\gamma}{f+(1-f)\eta_\gamma},
\end{equation}
where $f=f_{b,k}/f_{b,\gamma}$. It can be seen that when jet angle
evolution is considered, it is equivalent to increasing the $E_k$ by
a factor of $f$. For a double-sided jet,
$f_{b,\gamma}=1-cos(\theta_{\gamma})$, and
$f_{b,k}=1-cos(\theta_{k})$, where $\theta_{\gamma}$ and
$\theta_{k}$ are the opening angles during the prompt and the
afterglow phases, respectively.

The hydrodynamic evolution of the jet has been studied by a number
of authors (e.g., Rhoads 1997, 1999; Panaitescu \& M\'{e}sz\'{a}ros
1999; Sari, Piran, \& Halpern 1999; Moderski, Sikora, \& Bulik 2000;
Huang et al. 2000; Kumar \& Panaitescu 2000). We adopt the evolution
equation group presented in Huang et al. (2000), which describes the
overall evolution of relativistic GRB ejecta from the an
ultra-relativistic phase to a non-relativistic phase. The equations
are quoted as follows:
\begin{equation}
\frac{dR}{dt}=\beta c\gamma(\gamma+\sqrt{\gamma^{2}-1}),
\end{equation}
\begin{equation}
\frac{dm}{dt}=2\pi R^{2}\beta c\gamma(\gamma+\sqrt{\gamma^{2}-1})(1-cos\theta)nm_{p},
\end{equation}
\begin{equation}
\frac{d\theta}{dt}=\frac{c_{s}(\gamma+\sqrt{\gamma^{2}-1})}{R},
\end{equation}
\begin{equation}
\frac{d\gamma}{dt}=-\frac{2\pi R^{2}\beta
c\gamma(\gamma+\sqrt{\gamma^{2}-1})(\gamma^{2}-1)(1-cos\theta)nm_{p}}{M_{ej}+\epsilon
m+2(1-\epsilon)\gamma m},
\end{equation}
where $R$ is the radial coordinate in the burst source frame,
$\gamma$ is the bulk Lorentz factor of the ejecta
[$\beta=(\gamma^{2}-1)^{1/2}/\gamma$], $m$ is the swept-up mass, $n$
is the number density of the surrounding interstellar medium (ISM),
$m_{p}$ is the mass of the proton, $\theta$ is the half-opening
angle of the jet,
$c_{s}=\{[(\hat{\gamma}(\hat{\gamma}-1)(\gamma-1))]^{1/2}/[1+\hat{\gamma}(\gamma-1)]^{1/2}\}c$
is the comoving sound speed, $\hat{\gamma}=(4\gamma+1)/(3\gamma)$ is
the adiabatic index, and $\epsilon$ is the radiation
efficiency.\footnote{This efficiency is the one during the
deceleration phase of ejecta after burst which is different from the
GRB efficiency} $M_{ej}$ is the ejecta mass that is defined by
$E_{iso}(1-cos\theta_{0})=\gamma_{0}M_{ej}c^{2}$, where $E_{iso}$,
$\theta_{0}$, and $\gamma_{0}$ are the burst energy (isotropic-
equivalent), the initial half-opening angle of jet, and the initial
Lorentz factor, respectively.

We first adopt typical parameters as follows in order to study the
jet sideways effect: $\gamma_{0}$=300, $E_{iso}$=10$^{53}$ ergs,
$\theta_{0}$=0.1 radians, and $n=1$ cm$^{-3}$. In our calculation,
we take $\theta_{\gamma}$ to be the initial half-opening angle of
the jet $\theta_{0}$, and $\theta_{k}$ as the one at 10 hr. With
these typical parameters, we derive the evolution of the jet opening
angle, which is shown in Figure 1. We find that $\theta_{k}\sim0.14$
at $\sim 10$ hr. Therefore, $f\sim 2$, and hence
$\eta_\gamma^c=\eta_\gamma/(2-\eta_\gamma)$. This indicates that the
lower the efficiency, the more significant the reduction is on the
efficiency from this effect. For a burst with $\eta_\gamma=50\%$,
$\eta_\gamma^c=2\eta_\gamma/3$, while for $\eta_\gamma=10\%$,
$\eta_\gamma^c=\eta_\gamma/2$. The initial parameters, $E_{iso}$ and
$n$, vary between bursts. Figure 1 also shows the evolution of the
jet opening angle for various values of $E_{iso}$ and $n$;
$\eta_\gamma^c$ as a function of the initial parameters is shown in
Figure 2. It is found that the correction to $\eta_\gamma$ tends to
be more significant for a burst with smaller $\theta_0$, lower
$E_{iso}$, and a denser ambient medium.

\begin{figure}
\begin{center}
\includegraphics[angle=0,scale=1]{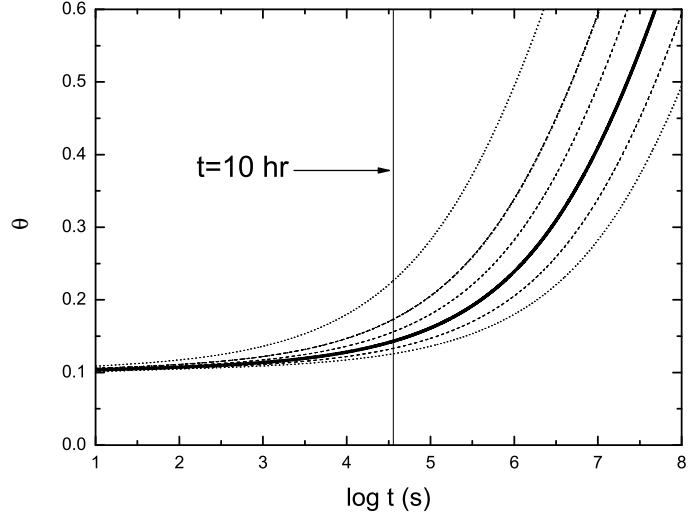}
\caption{Evolution of the jet half-opening angle. The vertical solid
line marks the position of t=10 hr. The thick solid line is for the
typical parameters. The dotted lines are for the typical parameters
with $n$ = 10,000, 100, and 0.01 cm $^{-3}$ from the top to the
bottom. The dashed lines are for the typical parameters with
$E_{iso}=10^{51}$, $10^{52}$, and $10^{54}$ ergs from the top to the
bottom. Note that the lines for $n=100$ cm$^{-3}$ and
$E_{iso}=10^{51}$ ergs are almost superposed.}
\end{center}
\end{figure}

\begin{figure}
\begin{center}
\includegraphics[angle=0,scale=1]{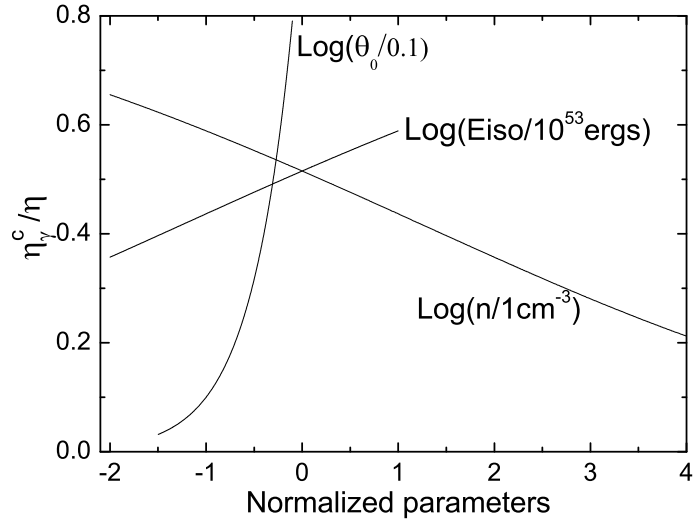}
\caption{The $\eta_\gamma^c$ in some extent of the initial
parameters, which is normalized to $\eta=0.1$ and the initial
parameters to the typical values. When one parameter varies, the
others are taken as the typical values.}
\end{center}
\end{figure}

\section{Case Studies}
We take two pre-\emph{Swift} bursts, GRB 980703 and GRB 000926, as
examples. The microphysics parameters of the two bursts have been
well constrained with high-quality broadband afterglow data (Yost et
al 2003). For calculating a more accurate hydrodynamic evolution,
the radiative loss during the early afterglow phase cannot be
neglected. Since the energy of the accelerated electrons behind the
blast wave is lost both through the synchrotron radiation (inverse
Compton cooling is neglected) and through the expansion of the
fireball, the radiative efficiency is the ratio of the synchrotron
power to the total power, namely (Dai \& Lu 1998,1999b)
\begin{equation}
\epsilon=\epsilon_{e}\frac{t^{\prime-1}_{syn}}{t^{\prime-1}_{syn}+t^{\prime-1}_{ex}},
\end{equation}
where $\epsilon_{e}$ is the fraction of shock energy given to the
electrons, $t^{\prime}_{syn}=6\pi
m_ec/(\sigma_{T}B^{\prime2}\gamma_{e,min})$ is the synchrotron
cooling time, with $\sigma_{T}$ the Thompson cross section and
$\gamma_{e,min}=\epsilon_{e}(\gamma-1)m_{p}(p-2)/m_{e}(p-1)+1$ the
minimum Lorentz factor of the random motion of electrons in the
comoving frame, and $t^{\prime}_{ex}=R/(\gamma c)$ is the comoving
frame expansion time. With the numerical factor derived by FP06, we
derive
\begin{equation}
E_{k}=10^{53}\textrm{ergs}~A(p)~R_{l}~
L_{X,46}^{4/p+2}(\frac{1+z}{2})^{(2-p)/(p+2)}\epsilon_{B,-2}^{-(p-2)/(p+2)}\epsilon_{e,-1}^{4(1-p)/(p+2)},
\end{equation}
where
\begin{equation}
A(p)=\{\frac{1.35\times10^{6}\times(1-p/2)}{0.42^{(2-3p)/4}~(7.6\times10^{11})^{(p-1)/2}~C_{p}^{(p-1)}~
[(2.42\times10^{18})^{(1-p/2)}-(4.84\times10^{16})^{(1-p/2)}]}\}^{\frac{4}{(p+2)}}.
\end{equation}
Here $p$ is the spectral index of the electron distribution, and
$C_{p}=13(p-2)/3(p-1)$, $R_{l}\sim
[t(10h)/T_{90}]^{17\epsilon_{e}/16}$; is a factor that accounts for
radiative losses during the first 10 hr following the prompt phase
(Sari 1997, LZ04), $L_{X,46}=L_{X}/10^{46}$  is the isotropic X-ray
afterglow luminosity at 10 hr (using the value given by Berger et
al. 2003), and $\epsilon_{B}$ is the fraction of shock energy given
to the magnetic field (the inverse Compton effect is neglected).
$E_{k}$ is the isotropic kinetic energy at $t=10$ hr.

Assuming the jet opening angle at 10 hour is $\theta_{10}$, the
initial kinetic energy of ejecta is $E_{0}=E_{k}(1-cos\theta_{10})$.
We thus obtain a boundary condition that when $t=t_{10}$,
$\theta=\theta_{10}$. In addition, in the standard model, there is a
light-curve break occurring at the time ($t_{j}$) when the bulk
Lorentz factor of the shock has slowed to $\gamma\sim\theta^{-1}$
(Rhoads 1997; Sari et al. 1999). This can be taken as another
boundary condition; namely, when $t=t_{j}$, $\gamma=\theta^{-1}$.
Note that the second condition may depend on the assumed initial
Lorentz factor. However, we find that the calculated evolutions of
the Lorentz factor with different assumed initial values (from 100
to 500) are almost the same during the late time ($\gtrsim$1000 s
with the typical parameters), which is understandable. At the late
time, the mass of the jet is dominated by the swept-up mass, so the
initial mass of ejecta, which is defined by the initial Lorentz
factor with a given initial kinetic energy, weakly affects the
evolution of the Lorentz factor of a jet. We adjust $\theta_{10}$
and $\theta_{0}$ in a proper range until the above two boundary
conditions are satisfied. Then the initial jet angle and the initial
kinetic energy are derived, and the efficiency can be calculated.
Our results are listed in Table 1. It can be seen that the corrected
efficiency is significantly lower than that derived by LZ04 and
FP06. Note that the $E_k$ derived in this analysis is different from
LZ04 and FP06 due to the different microphysics parameters used. If
we use the kinetic energy that we obtain, the efficiencies of GRB
980703 and GRB 000926 derived with equation (1) are 0.135 and 0.103,
respectively, while with equation (2), they are only 0.079 and
0.052, respectively.

\begin{table}[]
  \caption[]{ The parameters and the calculated GRB efficiency. $E_{\gamma}$ is the isotropic
  radiative energy from Bloom et al. (2003), microphysics parameters, $n$ and $t_{j}$
  from Yost et al. (2003). Here we
  assume $\gamma_{0}=300$. $a$ and $b$ are the
  efficiency from LZ04 and FP06, respectively.}
  \label{Tab:publ-works}
  \begin{center}
  \begin{tabular}{|c|c|c|c|c|c|c|c|c|c|c|}
  \hline
   &$E_{\gamma}$&&&&&E$_{k}$&$n$&$t_{j}$&$\eta_\gamma^c$\\
   GRB& $(10^{52}$ergs)&$\theta_{0}$&$\epsilon_{e}$&$\epsilon_{B}$&$p$&($10^{53}$ergs)&(cm$^{-3}$)&(days)&(a,b)\\
   \hline
 &&&&&&&&&0.08\\
980703&6.01&$8.37^{\circ}$&0.27&0.0018&2.54&3.85&28&3.4&(0.71,0.21)\\
 \hline
 &&&&&&&&&0.05\\
000926 & 27.97&4.59$^{\circ}$& 0.15&0.022&2.79& 24.3& 16&2.6&(0.74,0.23)\\

 \hline
\end{tabular}
  \end{center}
\end{table}

\section{Discussions and conclusion}
We have investigated the correction on the estimate of $\eta_\gamma$
by considering the jet sideways effect. Our numerical study shows
that for a stronger burst, a burst with a narrower jet opening
angle, or a burst in a more dense environment (Dai \& Lu 1999a),
this effect tends to be more significant. Using the typical
parameters, the corrected efficiency may be reduced by a factor of
$\sim 0.5$ for a GRB with typical parameters. We use two
pre-\emph{Swift} bursts, GRB 980703 and GRB 000926, as detailed case
studies, further confirming our numerical analysis results.

Several authors argued that the efficiencies of some {\em Swift}
bursts with long time X-ray flattening are as high as $\sim$75\%-
90\% (Zhang et al. 2006; Nousek et al. 2006; Ioka et al. 2006;
Granot et al. 2006) and that the internal-shock model for GRBs meets
the challenge of a so-called efficiency crisis (e.g., Ioka et al.
2006). As mentioned in \S 1, the estimate of the GRB efficiency is
strongly affected by many effects and depends on unobservable
physical parameters. As we show in this analysis, the small jet
sideways effect may result in a considerable reduction of the
efficiency. More recently, Zhang et al. (2007) estimated the kinetic
energy with the data of early afterglows, such as the deceleration
time (when the blast wave is decelerated). From the end of GRB to
the deceleration time, the jet sideways expansion is insignificant,
and thus this effect on the estimate of the GRB efficiency can be
negligible. However, they found that the derived GRB efficiency is
highly variable from a few percent to $>$90\% (at the deceleration
time). If these results are intrinsic but not caused by a parameter
selection effect or by the uncertainties of early afterglows, they
may refresh our understanding of the GRB phenomenon, and they may
require an improvement in the conventional shock model.

\begin{acknowledgements}
We thank Y. Z. Fan for helpful discussion and suggestions. We also
acknowledge the referee for constructive comments and suggestions.
This work is supported by the National Natural Science Foundation of
China (grants 10443003 and 10573030) and the Natural Science
Foundation of Yunnan (2003 A0025Q).
\end{acknowledgements}

\end{document}